\documentclass[letterpaper]{article}

\usepackage{natbib,alifeconf}  %% The order is important
\usepackage{url,hyperref}
\usepackage{amsmath,amssymb}
\usepackage{booktabs}
\usepackage{placeins}
\usepackage{dblfloatfix}

\graphicspath{{figures/}}

\title{Democracy on Rugged Landscapes: Phase Transitions in Optimal Voting Rules}

\author{
    Josh Nunley$^{1}$ \\
    \mbox{}\\
    $^1$Indiana University, Bloomington, IN, USA \\
    joshnunl@iu.edu
}

\begin{document}

\maketitle

\begin{abstract}
Laws and institutions shape individual outcomes through complex interactions with
citizens' diverse circumstances, yet how different voting methods navigate this
coupled landscape remains poorly understood. We model collective governance as
optimization on NK fitness landscapes, where shared bits (laws) are updated by
voting while individual bits (personal traits) remain fixed. A cross-dependency
parameter $\alpha$ controls how legislation's effects depend on individual
circumstances. We compare eight standard voting methods and a generalized
scoring family across landscape ruggedness $K \in \{1,\ldots,20\}$ and
$\alpha \in [0,1]$ with 1000 runs per configuration.

Under direct democracy, the optimal voting method undergoes sharp phase transitions as a
function of landscape complexity: cardinal score voting dominates on smooth
landscapes, ordinal scoring with $p=0.35$ at low-to-moderate
ruggedness, Borda count across a wide middle range, and STAR voting at the
highest complexity. A two-parameter empirical formula reduces the
$(K, \alpha)$ plane to a single complexity axis for visualization. Borda count
achieves the highest mean fitness and lowest variance across most of the
parameter space.

We further introduce a representative democracy model parameterized by identity
weight $\beta$ and candidate self-interest $p_{\mathrm{self}}$. Representation
reshapes the complexity-dependent structure even under favorable conditions: cardinal
score voting dominates across most regimes, with plurality emerging as the top
method at high $\beta$ and low-to-moderate $p_{\mathrm{self}}$.

\end{abstract}

\section{Introduction}
Democratic governance requires a population of individuals with diverse needs and
circumstances to make binding collective decisions (laws) that affect everyone.
A tax policy, zoning regulation, or healthcare mandate interacts differently with
each citizen depending on their occupation, health, location, and other personal
attributes. This interaction between shared institutions and individual
heterogeneity is a defining feature of law as a complex adaptive
system~\citep{ruhl1996fitness}, yet formal models of voting rarely account for
it.

We propose a framework that captures this structure directly. Each individual is
represented as a binary string on an NK fitness landscape~\citep{kauffman1993origins},
partitioned into a \emph{shared portion} (bits modified collectively through
voting, representing laws) and an \emph{individual portion} (bits that are
fixed and unique to each person, representing personal traits). The NK epistatic
structure couples these portions: the fitness contribution of each law-bit
depends on nearby trait-bits, and vice versa. A cross-dependency parameter
$\alpha$ controls the fraction of each bit's dependencies that cross between
portions. At $\alpha=0$, dependencies are purely internal: laws interact only
with other laws, so legislation affects all citizens identically. At $\alpha=1$,
dependencies are purely cross-portion: every law's effect depends on individual
traits, but traits lack complex internal interactions. Intermediate values produce
the richest landscapes. At $\alpha \approx 0.5$, both internal and
cross-dependencies are present, creating the highest effective complexity. This
parameter captures a simple spectrum: from purely public goods ($\alpha=0$) to
policies whose effect is entirely mediated by
personal circumstances ($\alpha=1$).

Within this framework, each round of voting is a step of collective optimization
on a rugged landscape. Eight standard voting methods (plurality, approval, score, Borda, IRV, STAR, minimax, and random dictator) aggregate individual preferences differently, and thus navigate the landscape differently over many rounds.

All voting methods receive the same input: a utility matrix
$U \in \mathbb{R}^{M \times 2^V}$ whose $(i,p)$ entry is the fitness voter~$i$
would obtain under proposal~$p$. A voting rule can therefore be viewed as a
\emph{transformation of the same utility field}: different methods preserve,
compress, or discard different aspects of $U$ before selecting a winner. For
example, plurality keeps only each voter's top choice, score voting sums raw
cardinal values, and Borda discards magnitudes but preserves full within-voter
ordinal structure. From this viewpoint, voting methods differ not only in their
standard social-choice properties but also in the kind of information processing
they perform.

Beyond direct democracy, we develop a unified model of \emph{representative
democracy} motivated by two simple distinctions in representation. The first is
whether a candidate's platform is delegate-like, reflecting constituency welfare,
or trustee-like, reflecting the candidate's own fitness in the model. We
parameterize this as $p_{\mathrm{self}} \in [0,1]$, the probability a
candidate's platform reflects their own interest rather than their constituency's
collective welfare. The second is \emph{identity versus policy voting}: do
citizens elect representatives who share their background and circumstances
(descriptive representation~\citep{mansbridge1999should}) or those whose
platforms promise the best outcomes? We parameterize this as $\beta \in [0,1]$, the weight on
identity-based versus policy-based candidate evaluation. Candidates' constituencies
are defined by nearest-neighbor Hamming distance in individual traits. The cross-dependency $\alpha$
then plays a natural role: when $\alpha=0$, individual traits are irrelevant to
policy outcomes and identity voting carries no information. When $\alpha=1$,
traits fully mediate policy effects, making identity a strong proxy for policy
alignment.

We study two questions. First, how does voting method choice affect long-run
collective welfare (mean fitness, variance, and distributional outcomes) across
varying landscape ruggedness $K$ and law-trait coupling $\alpha$? Second, how
does the introduction of representation, and the choice of $\beta$ and
$p_{\mathrm{self}}$, alter these dynamics, and in what parameter regimes does
representation help or hurt?

We use the term \emph{phase transition} in an empirical sense: as $K$ and $\alpha$ vary, the identity of the best-performing voting rule changes abruptly across narrow regions of parameter space, producing stable performance regimes separated by sharp boundaries. These are not thermodynamic phase transitions, but regime transitions in the optimizer induced by changes in landscape structure.

The central interpretation is that voting rules act as information transforms: they preserve some features of the voter-utility matrix while discarding others, and landscape complexity determines which preserved features are useful.

Our main finding is that voting methods undergo sharp phase transitions as a
function of landscape complexity: cardinal score voting dominates on smooth
landscapes, ordinal scoring with $p=0.35$ at low-to-moderate
ruggedness, Borda count across a wide middle range, and STAR voting at the
highest complexity. These transitions are
well-described by a two-parameter empirical formula that reduces the $(K, \alpha)$
plane to a single complexity axis for visualization. Borda count achieves both the highest mean fitness and lowest variance across the majority of the parameter space. Under representative democracy, the complexity-dependent
regime structure is reshaped even under favorable conditions: cardinal score voting
dominates across most regimes, with plurality emerging as the top method at
high $\beta$ and low-to-moderate $p_{\mathrm{self}}$.
These results also suggest a
new research question: whether voting rules can be ordered by computable
properties of the utility transformation they implement, and whether such
properties predict performance across environments.

\begin{figure}[t!]
  \centering
  \includegraphics[width=\columnwidth]{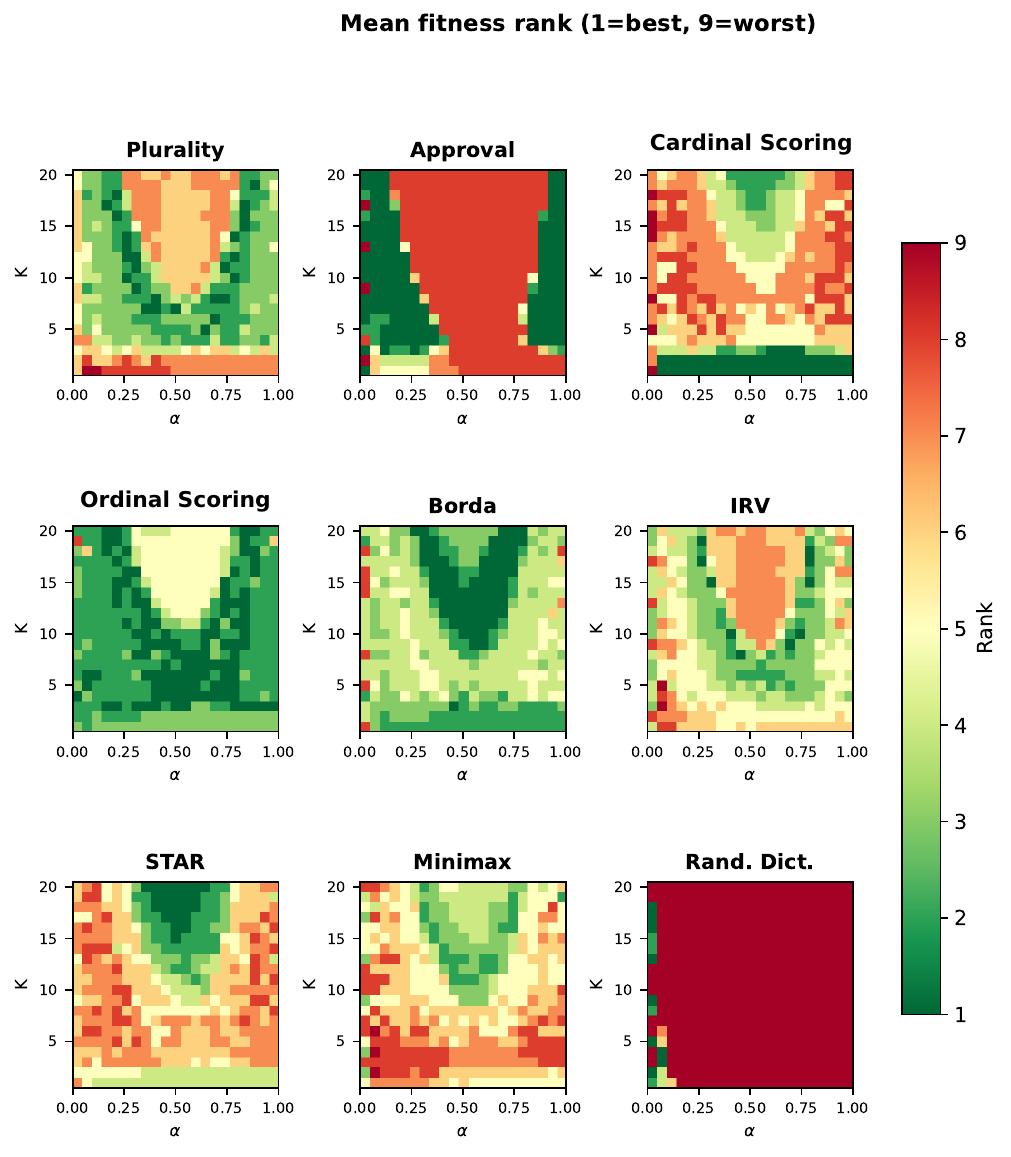}
  \caption{Mean fitness rank (1=best, 9=worst) for each voting method across the full $(K, \alpha)$ grid. Green indicates top performance; red indicates poor performance. Regime transitions are visible as sharp boundaries between green and red regions.}
  \label{fig:mean_rank}
\end{figure}

\section{Related Work}

\paragraph{NK landscapes and collective search.}
The NK landscape model~\citep{kauffman1993origins} provides a tunable framework
for studying optimization on rugged fitness landscapes, parameterized by the
number of components~$N$ and the degree of epistasis~$K$. It has been widely
applied to organizational decision-making~\citep{levinthal1997adaptation,
rivkin2000imitation}, where groups search for high-fitness configurations under
varying degrees of interdependence. In these models, organizations navigate
rugged landscapes through hierarchical decomposition, imitation, or parallel
search. Our work extends this tradition by modeling the search process as
democratic voting rather than managerial or evolutionary search, and by
introducing a partition between shared (law-like) and individual (trait-like)
components that creates voter heterogeneity absent from prior NK collective
search models.

\paragraph{Computational social choice.}
The formal study of voting methods has a long
history~\citep{arrow1951social, gibbard1973manipulation, satterthwaite1975strategy}. Arrow's impossibility
theorem established that no rank-order voting system can satisfy a small set of
reasonable criteria simultaneously, while the Gibbard--Satterthwaite theorem
showed that deterministic, non-dictatorial methods are susceptible to strategic manipulation. Computational approaches have
since analyzed the complexity, manipulability, and axiomatic properties of
various methods~\citep{brandt2016handbook}. However, most analyses focus on
single-shot elections or equilibrium outcomes rather than iterated collective
optimization over many rounds, which is the regime most relevant to ongoing
governance. Our framework bridges this gap by evaluating voting methods as
long-run optimizers on a landscape where the quality of collective decisions
accumulates over time.

\paragraph{Law as a complex adaptive system.}
Legal scholars have increasingly applied complexity theory to
law~\citep{ruhl1996fitness, ruhl2008complexity}, arguing that legal systems
exhibit emergent behavior, path dependence, and co-evolution with the populations
they govern. Our model operationalizes this perspective: laws (shared bits) and
individual traits co-determine fitness through epistatic interactions, and the
voting process drives adaptation of the legal component. The cross-dependency
parameter $\alpha$ formalizes the degree to which legislation interacts with
individual circumstances, a central concern in regulatory design that
complexity-theoretic analyses of law have identified but not previously modeled
at the level of voting mechanisms.

\section{Model}
We model a population of $M$ individuals making collective decisions on an NK
fitness landscape. Each individual~$i$ is represented by a binary string
$\mathbf{x}_i \in \{0,1\}^N$, partitioned into a \emph{shared portion}
$\mathbf{s} \in \{0,1\}^{N_s}$ (identical across all individuals, representing
laws) and an \emph{individual portion} $\mathbf{t}_i \in \{0,1\}^{N_t}$ (unique
and fixed, representing personal traits), where $N_s + N_t = N$.

\subsection{Fitness Landscape}

Fitness is computed according to the standard NK
model~\citep{kauffman1993origins}. Each bit position $j$ contributes a fitness
value $f_j$ that depends on the bit at position~$j$ and $K$ other positions
determined by a dependency matrix~$D$:
\begin{equation}
  F(\mathbf{x}_i) = \sum_{j=1}^{N} f_j\!\left(\mathbf{x}_i[j],\;
    \mathbf{x}_i[d_{j,1}],\ldots,\mathbf{x}_i[d_{j,K}]\right),
\end{equation}
where $d_{j,1},\ldots,d_{j,K}$ are the $K$ positions on which position~$j$
depends. Each $f_j$ maps a $(K\!+\!1)$-bit substring to a fitness value drawn
uniformly from $[-1, 1]$. Increasing $K$ increases the ruggedness of the
landscape, creating more local optima and making optimization harder.

\subsection{Cross-Dependency Parameter $\alpha$}

We control the coupling between shared and individual portions through a
cross-dependency fraction $\alpha \in [0,1]$. For each bit position~$j$, a
fraction~$\alpha$ of its $K$ dependencies are drawn from the \emph{other}
portion, while the remaining $1-\alpha$ are drawn from the same portion.
At $\alpha=0$, dependencies are entirely internal: laws affect all citizens
identically. At $\alpha=1$, dependencies are entirely cross-portion: individual
traits mediate all policy effects but lack complex internal structure. At
$\alpha \approx 0.5$, both are present, creating the highest effective
complexity.

\begin{figure}[t!]
  \centering
  \includegraphics[width=\columnwidth]{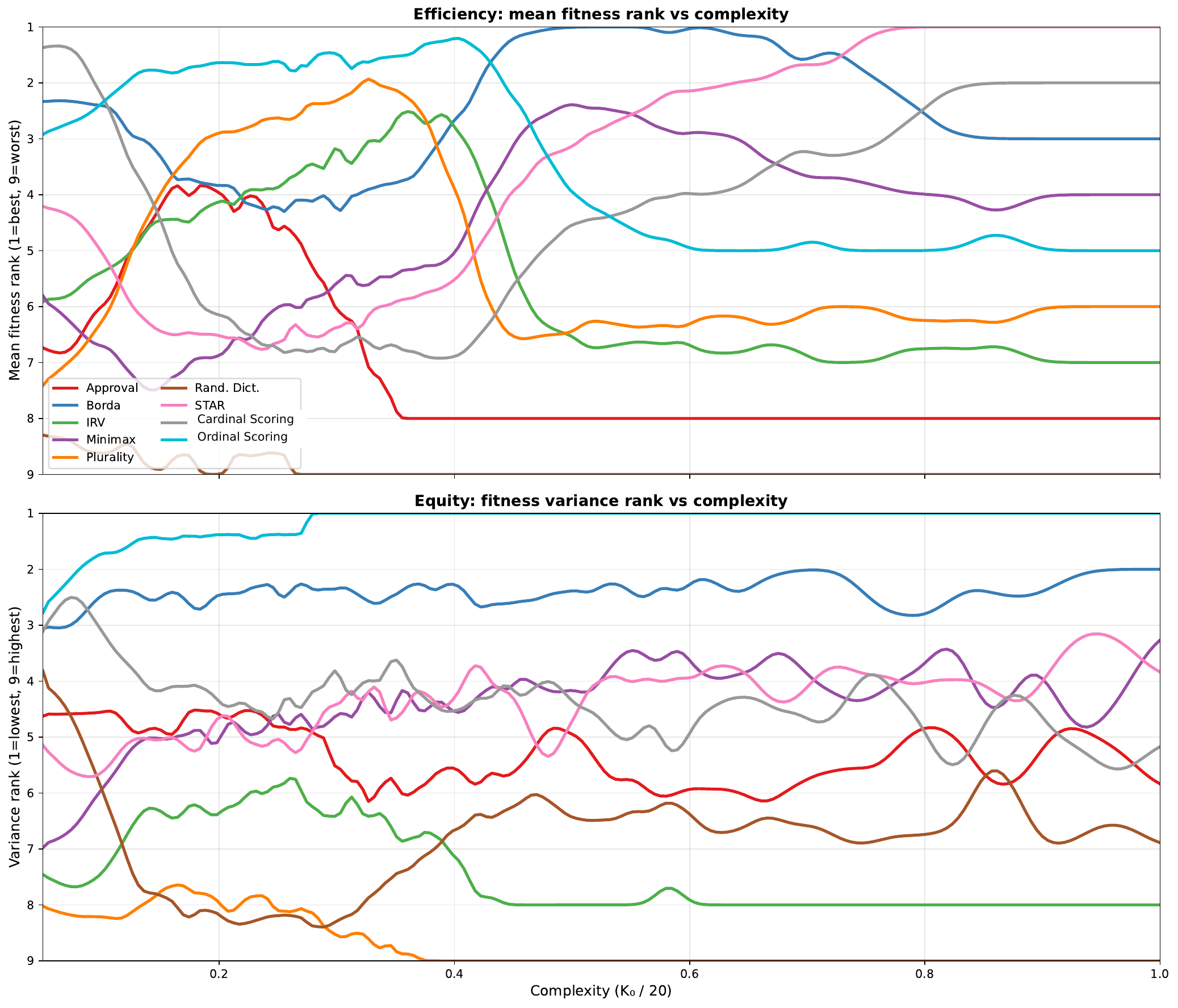}
  \caption{Mean fitness rank (top) and variance rank (bottom) vs.\ fitted complexity $K_0/20$. The progression Cardinal score $\to$ Ordinal scoring ($p\!=\!0.35$) $\to$ Borda ($p\!=\!1$) $\to$ STAR is visible for efficiency; both Borda and ordinal scoring ($p\!=\!0.35$) rank best for equity.}
  \label{fig:rank_vs_complexity}
\end{figure}

\subsection{Voting Process}

At each time step, the system selects $V$ bit positions uniformly at random from
the shared portion (we use $V=2$ in direct democracy unless otherwise noted).
This defines $2^V$ possible
proposals (all binary configurations of those positions). Each individual~$i$
evaluates all proposals by computing the fitness they would obtain under each
one, producing a utility matrix $U \in \mathbb{R}^{M \times 2^V}$.

A voting method $\mathcal{V}$ then acts on this matrix to select a winning
proposal, which is applied to all individuals' shared portions. Equivalently, we
may view each method as a function
\begin{equation}
  \mathcal{V}: \mathbb{R}^{M \times 2^V} \to \{1,\ldots,2^V\},
\end{equation}
or as a two-stage map that first transforms $U$ into proposal scores and then
selects the maximiser. Under this interpretation, all methods solve the same
task from the same input, but they differ in which information from $U$ they
retain. Some use only each voter's top choice, others use full ordinal
rankings, others use cardinal magnitudes, and pairwise methods reprocess the
same profile through head-to-head comparisons. We study eight standard methods
plus one member of a generalized scoring family:

\begin{itemize}
  \item \textbf{Plurality}: Each voter casts one vote for their highest-utility
    proposal; the proposal with the most votes wins.
  \item \textbf{Approval}: Each voter approves all proposals that improve their
    fitness over the status quo (or all tied-for-best if none improve); the most
    approved proposal wins.
  \item \textbf{Cardinal score}: Sum each voter's raw utility for each proposal; highest
    total wins.
  \item \textbf{Borda}: Each voter ranks proposals; points are assigned by rank
    and summed.
  \item \textbf{Instant-Runoff (IRV)}: Iteratively eliminate the proposal with
    the fewest first-place votes, redistributing those votes, until a majority
    winner emerges.
  \item \textbf{STAR}: Sum scores to find the top two proposals, then elect the
    one preferred head-to-head by more voters.
  \item \textbf{Minimax (Condorcet)}: Select the proposal whose worst pairwise
    margin of defeat is minimized.
  \item \textbf{Random Dictator}: A uniformly random voter's top choice is
    selected. Serves as a baseline.
  \item \textbf{Ordinal scoring ($p\!=\!0.35$)}: A positional scoring rule where rank~$k$
    (out of $n$) receives score $((n\!-\!1\!-\!k)/(n\!-\!1))^{0.35}$. This is a
    member of the family of positional scoring rules \citep{saari1995geometry}
    that interpolates between
    antiplurality ($p \to 0$) and plurality ($p \to \infty$), with Borda at
    $p=1$. At $p=0.35$, the rule is flatter than Borda, assigning substantial
    weight across the ranking while remaining purely ordinal. This exponent was
    identified from a sweep over the scoring family as the value that dominates
    at low-to-moderate complexity.
\end{itemize}

\subsection{Representative Democracy}

The representative model adds a second information bottleneck before aggregation occurs. Voters no longer vote directly over policy proposals; instead, they choose among candidates whose platforms may reflect either constituency welfare or personal self-interest. The parameter $\beta$ controls how much voters choose candidates by identity similarity rather than policy consequences, while $p_{\mathrm{self}}$ controls how often candidates choose platforms for themselves rather than their constituents.

In the representative variant, $C$ candidates are drawn uniformly at random
from the population. Each candidate $c$ represents a \emph{constituency}:
voters assigned by nearest-neighbor Hamming distance on individual traits.

\paragraph{Platform formation ($p_{\mathrm{self}}$).}
With probability $p_{\mathrm{self}}$, candidate $c$ follows a
\emph{trustee-like} rule and selects the proposal maximizing their own fitness;
with probability $1 - p_{\mathrm{self}}$, they follow a \emph{delegate-like}
rule and maximize mean constituency welfare. This setup interpolates between
constituency-focused ($p_{\mathrm{self}}=0$) and candidate-focused
($p_{\mathrm{self}}=1$) platform formation.

\paragraph{Candidate election ($\beta$).}
Voters evaluate candidates by a convex combination of \emph{policy utility}
$u_{\mathrm{policy}}(i,c) = F_i(\pi_c)$ and \emph{identity utility}
$u_{\mathrm{id}}(i,c) = 1 - d_H(\mathbf{t}_i, \mathbf{t}_c)/N_t$:
\begin{equation}
  u(i, c) = (1-\beta)\,\tilde{u}_{\mathrm{policy}}(i, c)
           + \beta\,u_{\mathrm{id}}(i, c),
\end{equation}
where $\beta \in [0,1]$ is the \emph{identity weight} and both components are
normalized to $[0,1]$. The same voting method used in direct democracy then
elects a winning candidate whose platform is implemented.

The $(\beta, p_{\mathrm{self}})$ plane nests four limiting cases.
Rows: election type (policy vs.\ identity) $\times$ platform type
(constituency vs.\ self-interested):

\begin{center}
\begin{tabular}{lll}
\toprule
Model & $\beta$ & $p_{\mathrm{self}}$ \\
\midrule
Trustee-like        & 0 & 1 \\
Delegate-like       & 0 & 0 \\
Identity--trustee-like  & 1 & 1 \\
Identity--delegate-like & 1 & 0 \\
\bottomrule
\end{tabular}
\end{center}

The cross-dependency $\alpha$ interacts with $\beta$: when $\alpha=0$, identity
utility carries no policy information and $\beta > 0$ adds variation unrelated to policy outcomes;
when $\alpha=1$, traits fully mediate policy effects, making identity a strong
proxy for policy alignment.

\section{Experiments}
All experiments use $N=50$ bit positions with a voting portion of $0.5$
($N_s = N_t = 25$), a population of $M=100$ individuals, $V=2$ bits voted on per
round, and all individuals sharing identical initial voting bits with unique,
randomly generated individual bits.

\subsection{Experiment 1: $K \times \alpha$ Sweep (Direct Democracy)}

We sweep landscape ruggedness $K \in \{1, 2, \ldots, 20\}$ and cross-dependency
$\alpha \in \{0.00, 0.05, 0.10, \ldots, 1.00\}$ (21 levels) under direct
democracy for all eight standard voting methods. Each configuration is run for
$150 + 50K$ iterations with 1000 independent runs, providing high-resolution
coverage of the full $(K, \alpha)$ plane. For each run we record mean fitness,
fitness variance, minimum and maximum fitness at every iteration.

In addition, we sweep a generalized positional scoring family parameterized by
exponent $p$: rank~$k$ (out of $n$) receives score $((n\!-\!1\!-\!k)/(n\!-\!1))^p$.
This family includes antiplurality ($p \to 0$), Borda ($p=1$), and plurality
($p \to \infty$) as special cases. We evaluate 28 exponents spanning
$p \in [0.05, 10]$ across the full $(K, \alpha)$ grid with 1000 runs per
configuration, identifying $p = 0.35$ as the exponent that dominates at
low-to-moderate complexity. This ninth method is included in all subsequent
analyses.

\subsection{Experiment 2: Representative Democracy}

\label{sec:exp2}

Using the representative democracy model described in Section~3, we sweep
$\beta \in \{0,\, 0.5,\, 1\}$ and $p_{\mathrm{self}} \in \{0,\, 0.5,\, 1\}$
(9 configurations) with $C=5$ candidates selected uniformly at random, $V=4$
bits voted on per round (16 proposals), across the full landscape grid:
$K \in \{1, \ldots, 20\}$ and $\alpha \in \{0.0, 0.1, \ldots, 1.0\}$.
All nine voting methods are tested, yielding 17{,}820 configurations with 500
independent runs each. The larger proposal space ($V=4$ versus $V=2$ in
Experiment~1) ensures that candidates can hold distinct platforms and
constituency welfare can meaningfully diverge from population welfare. Using
the same $(K, \alpha)$ grid as Experiment~1 allows direct comparison of the
fitted complexity axis under representation.

\section{Results}

\subsection{Direct Democracy: $K \times \alpha$ Sweep}

We present three results: (1)~method performance is regime-dependent, (2)~the $(K, \alpha)$ plane reduces to a single complexity axis via a fitted formula, and (3)~efficiency (mean fitness) and equity (fitness variance) dissociate.

\subsubsection{Mean Fitness Rank Phase Diagram}

Figure~\ref{fig:mean_rank} shows the rank of each voting method (1=best, 9=worst) by terminal mean fitness at every $(K, \alpha)$ configuration.

\textbf{No single method dominates.} The $(K, \alpha)$ plane is partitioned into regimes where different methods are optimal. At the lowest complexity ($K \leq 3$, $\alpha > 0$), cardinal score voting ranks first: a simple utilitarian sum suffices when the landscape is nearly additive. Among the eight classical methods, as $K$ increases to $4$--$9$, plurality takes over in the mid-$\alpha$ range. However, ordinal scoring with $p=0.35$ (identified from a sweep over exponents in $[0.05, 10]$) dominates this entire low-to-moderate regime, outperforming plurality while maintaining Borda's favorable equity properties (Figure~\ref{fig:rank_vs_complexity}). At moderate complexity ($K \approx 8$--$14$), Borda count dominates: full ordinal information becomes necessary for top performance on a rugged landscape. At the highest complexity ($K \geq 15$), STAR voting emerges as rank~1 in the mid-$\alpha$ region, with its two-stage mechanism (scoring followed by pairwise runoff) suited to highly rugged landscapes.

\textbf{Approval voting performs poorly across much of the mid-$\alpha$ region.} It exhibits a U-shaped pattern: it ranks first at extreme $\alpha$ values ($\alpha \approx 0$ and $\alpha \approx 1$) but drops to rank~7 across the entire mid-$\alpha$ band ($0.25 \leq \alpha \leq 0.80$) for $K \geq 5$. This boundary is sharp, not gradual. One plausible mechanism is that approval is thresholded against the status quo: at intermediate cross-dependency, the status quo becomes a weak proxy for the collective optimum, so the approval profile discards useful preference information.

\textbf{Random dictator is rank~9 except at $\alpha = 0$.} This confirms its role as a coordination-free baseline.

\subsubsection{A Fitted Complexity Parameter}

To visualize regime transitions on a single axis, we fit parabolic iso-complexity curves
\begin{equation}
  K = a\,(K_0 - 1)^2 \left(\alpha - \tfrac{1}{2} - \frac{b}{K_0 - 1}\right)^{\!2} + K_0,
  \label{eq:complexity}
\end{equation}
where $K_0 \in [1, 20]$ indexes complexity and $a$, $b$ are fitted by maximizing rank-vector uniformity along iso-$K_0$ contours (Gaussian-kernel-weighted rank variance, bandwidth $\varepsilon = 0.5$). The fitted values are $a = 2.35$, $b = 0.29$. The center $\mu(K_0) = \tfrac{1}{2} + b/(K_0-1)$ converges to $\tfrac{1}{2}$ at high complexity, consistent with $\alpha = \tfrac{1}{2}$ producing the richest dependency structure.
For $K_0 = 1$, where the parameterization is singular, we treat the degenerate
iso-complexity contour separately as the flat line $K=1$.

\subsubsection{Efficiency and Equity vs.\ Complexity}

Figure~\ref{fig:rank_vs_complexity} shows mean fitness rank (top) and variance rank (bottom) for all nine methods plotted against normalized complexity $K_0/20$. The regime transitions collapse cleanly onto this single axis for both measures.

\textbf{Efficiency.} Cardinal score dominates at low complexity, then ordinal scoring ($p\!=\!0.35$) takes over through the low-to-moderate range, Borda dominates the mid-range, and STAR emerges at the highest complexity. Among classical methods alone, plurality briefly leads in the low-to-moderate range, but $p\!=\!0.35$ strictly dominates it there.

\textbf{Equity.} Both Borda and ordinal scoring ($p\!=\!0.35$) achieve rank~1--2 for variance (most equitable) across nearly the entire complexity range. Ordinal scoring ($p\!=\!0.35$) achieves the best or near-best equity while simultaneously dominating efficiency at low-to-moderate complexity, a combination no classical method achieves. Plurality and random dictator consistently produce high variance; IRV shows poor equity at high complexity.

\begin{figure*}[t!]
  \centering
  \includegraphics[width=0.95\textwidth]{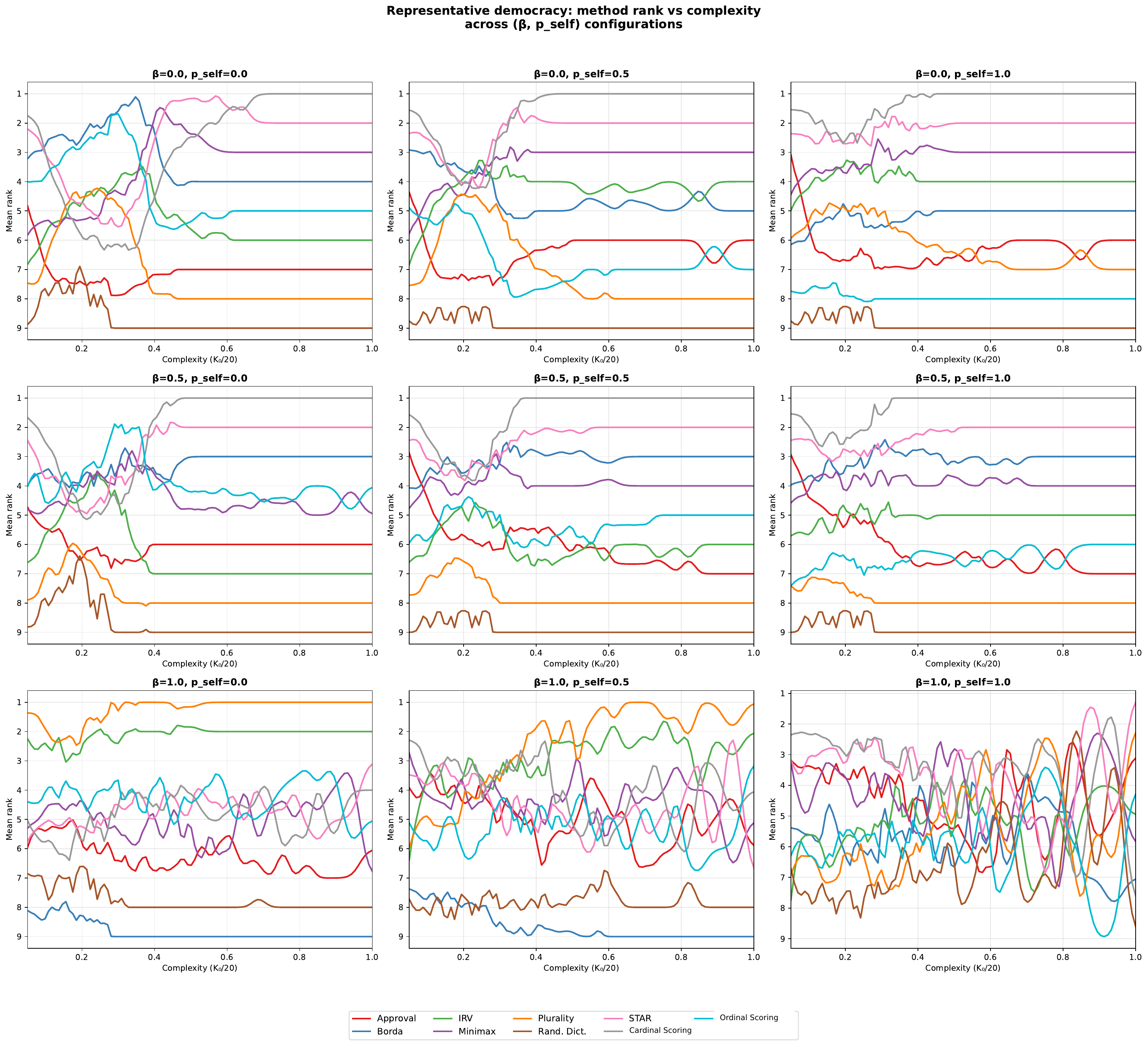}
  \caption{Mean rank vs.\ complexity for all nine methods across the $3 \times 3$ $(\beta, p_{\mathrm{self}})$ grid. Rows vary $\beta$ (identity weight: 0, 0.5, 1); columns vary $p_{\mathrm{self}}$ (candidate self-interest: 0, 0.5, 1). Representation reshapes the regime structure: high $p_{\mathrm{self}}$ produces chaotic rankings; high $\beta$ favors simpler methods.}
  \label{fig:rep_grid}
\end{figure*}

\subsection{Representative Democracy}

We sweep $\beta \in \{0, 0.5, 1\}$ and $p_{\mathrm{self}} \in \{0, 0.5, 1\}$ across the full $(K, \alpha)$ grid ($K \in \{1, \ldots, 20\}$, $\alpha \in \{0.0, 0.1, \ldots, 1.0\}$) with $V = 4$ bits, $C = 5$ candidates, and 500 runs per configuration.

\subsubsection{The $(\beta, p_{\mathrm{self}})$ Grid}

Figure~\ref{fig:rep_grid} presents a $3 \times 3$ grid of rank-versus-complexity curves, one panel for each $(\beta, p_{\mathrm{self}})$ configuration, using the same fitted complexity axis (Eq.~\ref{eq:complexity}) for comparability with direct democracy.

Cardinal score voting ranks first or near-first in most panels of the grid. The exception is the high-$\beta$, low-to-moderate-$p_{\mathrm{self}}$ regime (bottom-left two panels): plurality emerges as the top-ranked method here, since its concentrated signal resists the noise introduced by identity voting better than methods that require richer preference information.

\textbf{Regime structure persists under ideal representation.} At $(\beta = 0,\, p_{\mathrm{self}} = 0)$ (policy-based voting with delegate-like platform formation) the rank-versus-complexity curves most closely resemble direct democracy. Multiple methods compete across the complexity axis, and the progression from simple to sophisticated methods is visible. This configuration represents ideal representation in the model: voters evaluate candidates on policy merit, and candidates choose platforms for their constituencies.

\textbf{Rank ordering becomes chaotic as $p_{\mathrm{self}}$ increases.} Moving rightward in the grid, the rank curves grow increasingly chaotic. At $p_{\mathrm{self}} = 1$, no method clearly dominates at any complexity level; the rank curves cross repeatedly and erratically, with no stable ordering. Self-interested representatives introduce noise that overwhelms the complexity-dependent signal: the proposals implemented reflect individual self-interest rather than any aggregation of collective preferences, rendering the choice of voting method effectively random.

\textbf{Method differentiation decreases as $\beta$ increases.} The effect is less severe than under high $p_{\mathrm{self}}$. At $\beta = 1$, voters elect representatives based on shared traits rather than policy outcomes. The resulting representatives may share their constituents' circumstances but do not necessarily champion their preferred policies, reducing the signal available to the voting method.

\textbf{Voting method has little systematic effect at $(\beta = 1,\, p_{\mathrm{self}} = 1)$.} This corner (identity voting with candidate-focused, trustee-like platform formation) produces the least differentiated rank curves. Representatives are elected for who they are, not what they propose, and then implement whatever benefits themselves. No voting method, in this model, recovers a strong aggregation signal when the representation layer discards both policy information (high $\beta$) and constituency alignment (high $p_{\mathrm{self}}$).

\subsubsection{Delegate-like Advantage}

The delegate-like endpoint ($p_{\mathrm{self}} = 0$) consistently outperforms the candidate-focused, trustee-like endpoint ($p_{\mathrm{self}} = 1$) in mean fitness when $\alpha > 0$, with the largest gaps at mid-$\alpha$ and high $K$. At $K = 10$, $\alpha = 0.5$, the delegate-like advantage is 1.39 mean fitness units (best-method fitness of 3.45 vs.\ 2.04). At $K = 20$, $\alpha = 0.5$, the delegate-like endpoint more than doubles collective welfare (1.82 vs.\ 0.78). At $\alpha = 0$, the advantage vanishes entirely; on landscapes with no cross-dependency, self-interest and constituency welfare coincide.

\subsubsection{The $\beta$--$\alpha$ Interaction}

Identity-based voting ($\beta = 1$) underperforms policy-based voting ($\beta = 0$) most severely at mid-$\alpha$. At $K = 10$, $\alpha = 0.5$, the penalty is 1.57 fitness units (best-method fitness of 3.45 under $\beta = 0$ vs.\ 1.88 under $\beta = 1$). At $\alpha = 0$, the three $\beta$ values converge exactly: individual traits carry no policy information, so identity voting introduces noise but does not systematically bias outcomes. At $\alpha = 1$, the $\beta$ values nearly converge: traits fully mediate policy effects, making identity similarity a reasonable proxy for policy alignment.

\section{Discussion}

\subsection{Direct Democracy: Complexity-Dependent Optimization}

Under direct democracy, no voting method universally dominates: the optimal method depends on the complexity of the underlying decision landscape. Our framework asks a different question than the classical impossibility theorems of Arrow and Gibbard-Satterthwaite, which characterize methods by axiomatic properties on fixed preference profiles. We find that the \emph{empirical} performance ordering of methods is a function of landscape complexity and undergoes sharp phase transitions.

The progression Cardinal score $\to$ Ordinal scoring ($p\!=\!0.35$) $\to$ Borda ($p\!=\!1$) $\to$ STAR can be interpreted in terms of information aggregation depth. On smooth landscapes, preferences are nearly aligned and a simple sum of utilities suffices. As ruggedness increases, preferences fragment: a flatter ordinal scoring rule ($p\!=\!0.35$), which spreads weight across the ranking, outperforms both the raw sum and plurality's hard cutoff. At moderate complexity, full ordinal information becomes necessary for top performance (Borda, $p\!=\!1$), and at the highest complexity, even ordinal information must be combined with pairwise verification (STAR). Each successive method in the progression extracts richer information from voters' preferences, at increasing computational and communicative cost. Among classical methods, plurality occupies part of the regime where $p\!=\!0.35$ dominates, suggesting that plurality captures a real low-complexity signal but that its hard cutoff is too coarse to exploit it fully.

\subsection{Voting Rules as Utility Transforms}

In our model, all voting methods operate on the same utility matrix
$U \in \mathbb{R}^{M \times n}$ and can be understood as different
transformations of this common input: plurality compresses each row to an
argmax, Borda replaces magnitudes with ordinal ranks, score voting sums
cardinal values directly, and pairwise methods reprocess the profile through
head-to-head comparisons. Each transformation preserves different features of
$U$---top-choice information, ordinal structure, cardinal intensity, or pairwise
margins---while discarding others.

This framing suggests a research direction. Given assumptions about
the fitness landscape (ruggedness, cross-dependency structure), one might prove
which properties of the utility transform lead to optimal collective
optimization, connecting computational social choice to optimization theory.
For instance, our results suggest that on smooth landscapes, preserving cardinal
magnitudes is optimal (score voting), while on rugged landscapes, preserving
full ordinal structure (Borda) or pairwise comparisons (STAR) becomes necessary.
A formal theory characterizing this relationship could yield optimality results
for voting methods as a function of landscape properties, rather than relying on
axiomatic criteria alone.

\subsection{Borda's Robustness and Equity}

Under direct democracy, Borda count is the most robust classical method across the middle complexity range, combining high mean fitness with consistently low variance. Its worst observed rank is approximately 4. This complements classical axiomatic work that identifies Borda by distinctive structural properties \citep{young1974axiom}, but in a dynamic, multi-round optimization setting far from the one-shot environments usually studied in social choice theory.

\subsection{Representation as an Information Bottleneck}

Subject to the model's simplifications (sincere voters, binary candidate platforms, random NK landscapes; see Limitations), representation reshapes the regime structure even under favorable conditions. At $(\beta = 0,\, p_{\mathrm{self}} = 0)$, the direct democracy ordering is partially visible but already altered: representation acts as a lossy compression, and even candidates using the delegate-like rule cannot fully reconstruct the population-level signal that direct democracy exploits. The two representation parameters introduce qualitatively different distortions: high $p_{\mathrm{self}}$ produces chaotic rank curves with no stable method ordering, while high $\beta$ favors simpler methods.

Plurality performs comparatively well when voters select on identity rather than policy, since its concentrated signal is more resistant to noise than methods requiring richer preference information. Voting method choice matters most under delegate-like representation. As either $\beta$ or $p_{\mathrm{self}}$ increases, the binding constraint shifts from aggregation quality to representation quality.

Within these limits, the model identifies two ways the representation layer can
weaken the signal available to a voting rule. High $\beta$ captures selection on
identity rather than policy. High $p_{\mathrm{self}}$ captures candidates choosing
platforms for themselves rather than constituents. In both cases, voting rules
matter most when voters receive useful policy signals and candidates remain
aligned with constituents \citep{mill1861representative, achen2016democracy}.

\subsection{Limitations}

Several simplifications limit generalizability. All agents vote sincerely; strategic voting could qualitatively change method comparisons, particularly for IRV and Borda \citep{gibbard1973manipulation}. Individual portions are fixed; in reality, personal circumstances respond to collective decisions. The NK landscape is random; results are averages over realizations, not predictions for specific policy domains. The proposal space ($V = 2$ for direct, $V = 4$ for representative) is small. Our constituency assignment uses Hamming distance, a simplification of real representational geography. Finally, we study only three discrete values of $\beta$ and $p_{\mathrm{self}}$; finer resolution might reveal additional structure.

\section{Conclusion}

We have presented an NK landscape framework for studying voting methods as collective optimizers. The main findings are:

\begin{enumerate}
  \item \textbf{Complexity-dependent method ordering.} Under direct democracy,
    method performance undergoes sharp phase transitions: cardinal score $\to$
    ordinal scoring ($p\!=\!0.35$) $\to$ Borda ($p\!=\!1$) $\to$ STAR as complexity increases. No
    method universally dominates.

  \item \textbf{Borda's robustness.} Under direct democracy, Borda count is the most robust classical method across the middle complexity range, combining high mean fitness with consistently low variance.

  \item \textbf{A flatter ordinal scoring rule.} Under direct democracy, a sweep
    over the generalized scoring family reveals that $p=0.35$ (flatter than Borda) dominates plurality in the low-to-moderate regime while matching
    Borda's favorable variance properties.

  \item \textbf{Voting rules as utility transforms.} The model frames voting
    methods as transformations of a common utility matrix, suggesting that formal
    connections between transform properties and landscape-dependent optimality
    may be possible.

  \item \textbf{Representation as information bottleneck.} Under representative
    democracy, the regime structure is reshaped by candidate self-interest and
    identity voting, with simpler methods gaining advantage as representation
    degrades.
\end{enumerate}

Future work could explore:

\begin{itemize}
  \item strategic voting;
  \item formal connections between utility transform properties and landscape-dependent optimality;
  \item fitness-biased candidate selection to model incumbency advantage;
  \item a third, socially transmitted ``belief'' portion of the bitstring, so that opinion dynamics on networks can interact with collective optimization.
\end{itemize}

Within the limits of the model, the broader implication is that voting reform cannot be evaluated independently of assumptions about the structure of the policy environment. If policy effects are smooth and broadly aligned, cardinal aggregation performs well. If effects are rugged and heterogeneous, ordinal and pairwise structure become more valuable. Voting rules are therefore not only procedures for choosing winners; in this model, they are adaptive information-processing mechanisms whose performance depends on the landscape they are asked to navigate.

\section{Acknowledgments}
This work used the Big Red 200 supercomputer at Indiana University, supported by
Lilly Endowment, Inc., through its support for the Indiana University Pervasive
Technology Institute. AI tools were used to assist with editing the manuscript and
assisting with code development. Multiple independent AI systems were employed, with
outputs cross-referenced to identify inconsistencies. The author reviewed all
AI-generated and AI-edited content, taking full responsibility for the final
work. All code was further validated through systematic testing and
visualization.

\footnotesize
\bibliography{references}
\bibliographystyle{apalike}

\end{document}